\documentclass[preprint,showpacs,aps,preprintnumbers,amsmath,amssymb,nofootinbib]{revtex4}
\usepackage{bbm}
\usepackage{amsfonts}
\usepackage{slashed}
\usepackage{subfigure}

\usepackage{array}

\usepackage{booktabs}
\usepackage{color, soul}
\usepackage{mathrsfs}
\usepackage{epsfig}
\usepackage{CJK}
\usepackage{graphicx}
\usepackage{dcolumn}
\usepackage{bm}
\usepackage{amsmath}
\usepackage{amssymb}
\usepackage{multirow}

\begin{document}

\title{Explanation of the ATLAS Z-peaked excess by squark pair production in the NMSSM}

\author{Junjie Cao$^{1,2}$, Liangliang Shang$^1$, Jin Min Yang$^3$, Yang Zhang$^3$}

\affiliation{ $^1$  Department of Physics,
        Henan Normal University, Xinxiang 453007, China \\
  $^2$ Department of Applied Physics, Xi'an Jiaotong University, Xi'an 710049, China \\
  $^3$ State Key Laboratory of Theoretical Physics,
      Institute of Theoretical Physics, Academia Sinica, Beijing 100190,
      China }

\begin{abstract}
The ATLAS collaboration recently reported a $3\sigma$ excess in the leptonic-$Z+jets+E_{T}^{miss}$ channel.
We intend to interpret this excess by squark pair production in the Next-to-Minimal Supersymmetric Standard Model (NMSSM).
The decay chain we employ is $\tilde{q} \to q \tilde{\chi}_2^0 \to q \tilde{\chi}_1^0 Z$, where $\tilde{\chi}_1^0$
and $\tilde{\chi}_2^0$ denote the lightest and the next-to-lightest neutralinos with singlino and bino as their
dominant components respectively. Our simulations indicate that after considering the constraints from the ATLAS
searches for $jets + E_{T}^{miss}$ signal the central value of the excess can be obtained for $m_{\tilde{q}}
\lesssim 1.2 {\rm TeV}$, and if the constraint from the CMS on-$Z$ search is further considered, more than 10 signal events
are still attainable for $m_{\tilde{q}} \lesssim 750 {\rm GeV}$. Compared with the interpretation by gluino pair production,
the squark explanation allows for a significantly wider range of $m_{\tilde{q}}$ as well as a less compressed SUSY mass spectrum.
We also show that the squark explanation will be readily tested at the initial stage of the 14 TeV LHC.
\end{abstract}

\maketitle
\newpage

\section{Introduction}

Since the discovery of the Higgs-like particle by the ATLAS and CMS collaborations at the Large Hadron Collider (LHC) in 2012\cite{Higgs-discovery}, the main task of the LHC program has shifted to the searches for new physics beyond the Standard Model (SM). These searches cover a wide range of possible signatures, notably various combinations of jets (with or without b-tagging), the missing transverse energy ($E_T^{miss}$) and/or leptons. In this direction, the ATLAS collaboration has recently reported an intriguing excess at $3\sigma$ significance in the leptonic-$Z + jets + E_{T}^{miss} $ channel\cite{ATLAS-Z-Excess}. Based on the full dataset in 2012, which corresponds to about $20.3fb^{-1}$ integrated luminosity at the 8 TeV LHC, the collaboration observed 29 events for the on-Z electron and muon pair channels in contrast with the expected SM background number $10.6\pm3.2$, and no excess over the SM background was observed in any other signal region (SR)\cite{ATLAS-Z-Excess}.

So far there have been several attempts to explain the excess by the production of new physics particles, which are needed to decay with a sizeable rate into jets as well as at
least one Z boson and one invisible particle \cite{Explanation-1,Explanation-2,Explanation-3,Explanation-4,Explanation-5,Explanation-6,Explanation-7,Explanation-8,Explanation-9,Explanation-10}.
In the context of supersymmetric  theories (SUSY), in order to provide enough events after the rather tight cuts employed in \cite{ATLAS-Z-Excess}, the production of
a pair of moderately light gluinos was usually utilized for the explanation\cite{Explanation-1,Explanation-3,Explanation-4,Explanation-5,Explanation-6,Explanation-9,Explanation-10}.
The key point in doing this is to choose a lightest SUSY particle (LSP) with relatively suppressed couplings to squarks, so that
the gluino prefers to decay first into the neutralino other than the LSP,
and the neutralino subsequently decays into the LSP plus a Z boson. In the Minimal Supersymmetric standard model (MSSM), a higgsino-dominated neutralino has
very weak couplings to light flavor squarks, so one can naturally imagine a higgsino-dominated LSP and assume the decay chain $\tilde{g} \to q \bar{q} \tilde{\chi}_i^0 \to
q \bar{q} Z \tilde{\chi}_{1}^0$ ($q$ represents a light flavor quark, and $\tilde{\chi}_i^0$ denotes a gaugino-dominated neutralino) to interpret the
excess\footnote{In this case, the higgsino-dominated LSP plays
the same role as the gravitino in the ATLAS report \cite{ATLAS-Z-Excess} to interpret the excess.}. For this case, we note that the measured dark matter relic density
is not easy to obtain if only the neutralino serves as the dark matter candidate (see for example, Fig.1 in \cite{Natural-SUSY}).
In the Next-to Minimal Supersymmetric Standard Model (NMSSM)\cite{NMSSM}, however,
a singlino-dominated neutralino also has the property, and meanwhile if it acts as the LSP, the correct relic density can be achieved
by multiple annihilation channels\cite{NMSSM,Cao-NMSSM-DM}. So in this work we are interested in the interpretation of the $Z$-excess in the NMSSM
with a singlino-dominated LSP.

In the framework of the NMSSM with a singlino-dominated LSP, the gluino pair production with the three-body decay $\tilde{g} \to q \bar{q} \tilde{\chi}_2^0 \to q \bar{q} Z \tilde{\chi}_{1}^0$
has been studied for the $Z$-excess in \cite{Explanation-3,Explanation-6}. These works indicated
that only with simple assumptions on the relevant model parameters can the NMSSM explain the excess quite well. Explicitly speaking, it was found that after considering
the constraints from the ATLAS searches for $jets + E_{T}^{miss}$ signal, the NMSSM can reproduce the central value of the excess, and even if one further considers
the constraint from the CMS search for the leptonic-$Z + jets + E_{T}^{miss} $ channel which observed no excess in all SRs,  the event number of the ATLAS
on-$Z$ signal can still reach 11, which is about $1.2 \sigma$ away from the measured central value\cite{Explanation-6}. Moreover, as illustrated in \cite{Explanation-3}
the gluino explanation can reproduce well various distributions of the excess presented by the ATLAS collaboration. Albeit these advantages, we still
think that it is necessary to seek for other explanations since in the gluino explanation, the gluino mass is limited in a narrow range, and at same time
the relevant sparticle mass spectrum must be rather compressed to escape the constraints (see Fig.2 of \cite{Explanation-6}). In this work, we consider squark pair production
as an explanation of the excess, and in order to compare it with the gluino explanation, we make similar assumptions on the model parameters to those of \cite{Explanation-6}.
We find that in the squark explanation the range of the squark mass can be significantly extended in comparison with the gluino explanation, and moreover
the relevant sparticle mass spectrum may become less compressed. We also find that, just like the gluino explanation, the distributions of the excess can also be reproduced well
in the squark explanation.

This work is organized as follows. In Section \ref{sec:model}, we briefly introduce our scenario for the excess. In Section \ref{sec:results}, we perform a comprehensive analysis
of the relevant parameter space and present the results of our simulations on the Z-peaked excess. In Section \ref{sec:compare}, we choose some representative
parameter points, and exhibit their predictions on various distributions of the excess in comparison with the corresponding data provided by the ATLAS collaboration.
In section \ref{sec:test}, we briefly discuss future test of our scenario at the 14 TeV LHC. At last, we draw our conclusions in Section \ref{sec:concl}.

\section{Our Scenario for the $Z$-excess} \label{sec:model}

As one of the most economical extensions of the MSSM, the NMSSM contains one gauge singlet Higgs superfield $\hat{S}$ in its matter content. The superpotential of the general NMSSM
is given by\cite{NMSSM, Franke:1995tc}
\begin{equation}
\label{shlsuperpotential}
W_{NMSSM}=W_{MSSM}+\lambda \hat{S} \hat{H}_u \cdot \hat{H}_d  +\xi_F \hat{S} +\frac{\mu^\prime}{2} \hat{S}^2 + \frac{\kappa}{3} \hat{S}^3,
\end{equation}
where $W_{MSSM}$ is the superpotential of the MSSM without the $\mu$ term, $\hat{H}_u$ and $\hat{H}_d$ are the $SU(2)_L$ doublet superfields, $\kappa$ and $\lambda$ are
dimensionless coefficients, $\xi_F$ parameterizes the tadpole term and $\mu^\prime$ is a supersymmetric mass.

In this framework, the fermionic component field of $\hat{S}$ which is usually called singlino $\tilde{S}$ will mix with the gauginos and the higgsinos of the MSSM
to form neutralinos. In the basis $(\psi_1\equiv-i \tilde{B}, \psi_2 \equiv  - i \tilde{W}^0,
\psi_3 \equiv \tilde{H}_u^0, \psi_4 \equiv \tilde{H}_d^0, \psi_5 \equiv \tilde{S})$, the corresponding mass matrix
is given by \cite{NMSSM}
\begin{eqnarray}
\cal{M} = \left( \begin{array}{ccccc}
  M_1 & 0 &  \frac{e v_u}{\sqrt{2} c_w} & -\frac{e v_d}{\sqrt{2} c_w} & 0  \\
  0& M_2 & -\frac{e v_u}{\sqrt{2} s_w} & \frac{e v_d}{\sqrt{2} s_w} & 0  \\
   \frac{e v_u}{\sqrt{2} c_w} & -\frac{e v_u}{\sqrt{2} s_w} & 0 & -\mu_{eff} & -\lambda v_d  \\
    -\frac{e v_d}{\sqrt{2} c_w} & \frac{e v_d}{\sqrt{2} s_w} &  -\mu_{eff} & 0& -\lambda v_u \\
    0 & 0 &  -\lambda v_d & -\lambda v_u & 2 \kappa s + \mu^\prime
   \end{array} \right),
\end{eqnarray}
where $M_1$ and $M_2$ are soft gaugino masses, $v_u= v\sin \beta$ and $v_d = v \cos \beta$ are vacuum expectation values (vev)
of the Higgs fields $H_u$ and $H_d$ respectively, $\mu_{eff} = \mu + \lambda s $ with $s$ denoting the vev of
the singlet scalar field $S$, and $c_w = \cos \theta_W$. This matrix can be diagonalized by a $5 \times 5$ unitary matrix $N$, and consequently
neutralinos as mass eigenstates are defined by
\begin{eqnarray}
\tilde{\chi}_i^0 = \sum_{j=1}^5 N_{i j} \psi_j, \nonumber
\end{eqnarray}
where the mass order $m_{\tilde{\chi}_1^0}<m_{\tilde{\chi}_2^0}<\cdots<m_{\tilde{\chi}_5^0}$
is assumed. Obviously, the matrix element $N_{ij}$ measures the size of the $\psi_j$ component in $\tilde{\chi}_i^0$ state, and for the singlino-dominated
and bino-dominated neutralinos, their masses are mainly determined by the combination $2 \kappa s + \mu^\prime$ and $M_1$ respectively.
Moreover, with the help of $N_{ij}$ one can get the interactions of the neutralinos.
As shown in \cite{Explanation-6}, the $\bar{q} \tilde{\chi}_i^0 \tilde{q}$  coupling with $q$ denoting a light flavor quark
is determined by the gaugino components of $\tilde{\chi}_i^0$, and the  $\tilde{\chi}_i^0 \tilde{\chi}_j^0 Z$ coupling is determined by
the higgsino components of the neutralinos. By contrast, the $\tilde{\chi}_i^0 \tilde{\chi}_j^0 h$ coupling with $h$ denoting the SM-like Higgs boson
depends on all components of the neutralinos, and there may exist cancelations among different contributions.
These characters are helpful to understand our explanation of the $Z$-excess.

\begin{figure}
\centering
\includegraphics[width=10cm]{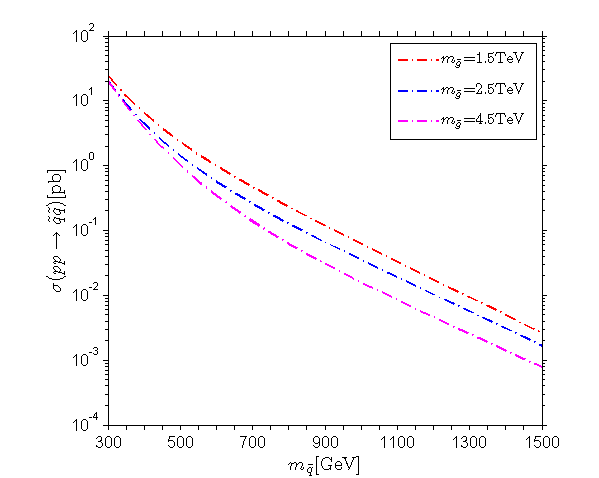}
\caption{The cross section of the squark pair production at the 8 TeV LHC for different gluino mass.}
\label{CrossSection}
\end{figure}

In the following, we intend to interpret the ATLAS on-Z excess by squark pair production. To make our explanation as simple as possible, we have the following assumptions:
\begin{itemize}
\item Only the first and second generation squarks are responsible for the excess. In our analysis, we assume a common mass $m_{\tilde{q}}$ for the squarks, then the cross
section for the squark pair production depends only on $m_{\tilde{q}}$ and $m_{\tilde{g}}$. We calculate the cross section at the NLO with the code Prospino \cite{prospino},
and show its dependence on $m_{\tilde{q}}$ at the 8 TeV LHC in Fig.\ref{CrossSection}.

\item The leptonic-$Z + jets + E_T^{miss}$ signal is generated by the cascade decay $\tilde{q}\to q \tilde{\chi}_2^0  \to q \tilde{\chi}_1^0 Z $. In order to maximize
this signal rate, we require both $Br(\tilde{q} \to q \tilde{\chi}_2^0)$ and $Br(\tilde{\chi}_2^0 \to \tilde{\chi}_1^0 Z)$ at roughly $100\%$, where the former requirement
can be satisfied if $\tilde{\chi}_1^0$ and $\tilde{\chi}_2^0$ are singlino-dominated and bino-dominated respectively, and only the two particles
in the neutralino and chargino sector are lighter than the squarks, and the latter condition can be realized if $ m_Z < m_{\tilde{\chi}_2^0} -
m_{\tilde{\chi}_1^0}  \leq m_h $ or if the $\tilde{\chi}_i^0 \tilde{\chi}_j^0 h$ interaction is significantly suppressed (see above
discussion and also our previous work \cite{Explanation-6}).

\item With the above assumptions, the parameters involved in our explanation are $m_{\tilde{q}}$, $m_{\tilde{g}}$, $\Delta m_1 \equiv m_{\tilde{q}} -  m_{\tilde{\chi}_2^0}$ and
$\Delta m_2 \equiv  m_{\tilde{\chi}_2^0} - m_{\tilde{\chi}_1^0}$.  In our discussion, we vary these parameters freely, but noting that  the process $pp\to \tilde{q}\tilde{q}
\to \tilde{\chi}_1^0 Z q \tilde{\chi}_1^0 Z q$ can also generate multi-jets $+ E_T^{miss}$ signal, we limit these parameters by the ATLAS searches for the multi-jets signal,
which were presented in \cite{ATLAS-6jet-new,ATLAS-6jet-old}. We also consider the CMS search for the leptonic-$Z + jets + E_T^{miss}$ signal \cite{CMS-Leptonic-Z} as an alternative
constraint on the parameters.
\end{itemize}

About our scenario for the $Z$-excess, we have following additional remarks:
\begin{itemize}
\item We ad hoc require that only $\tilde{\chi}_1^0$ and $\tilde{\chi}_2^0$ among the neutralinos are lighter than the squarks. This will simplify our analysis, but
on the other hand, since the rate and various kinematic distributions of the process  $pp\to \tilde{q}\tilde{q} \to \tilde{\chi}_1^0 Z q \tilde{\chi}_1^0 Z q$ are decided by
few parameters, the capability of our scenario to interpret the excess is limited given that the scenario must satisfy the constraints mentioned above. In fact, as
implied by the results of \cite{Explanation-8}, allowing the squarks to decay in multiple ways facilitates SUSY to balance the ATLAS signal and the constraints, and is
thus able to explain the excess in a better way.  This, however, needs an intensive scan over a higher dimensional SUSY parameter space, and for each parameter point,
simulations on various SUSY signals have to be done to compare with the corresponding experimental data. Such calculations
are very time-consuming, and are beyond the capability of our cluster.
\item Again for the sake of simplicity, we do not consider the effect of the third generation squarks in our analysis. These squarks have been tightly
limited by the SM-like Higgs boson mass, and are preferred to be heavy\cite{Higgs-Mass-1,Higgs-Mass-2,Higgs-Mass-3}. For some optimized points in Fig.\ref{fig2} for
the $Z$-excess, we once included their contributions to the leptonic-$Z + jets + E_T^{miss}$ and the multi-jets $+ E_T^{miss}$ signals by assuming the degeneracy
of all squarks. However, we did not find any improvement on our explanation due to the constraints we considered.
\item The assumptions on the properties of the LSP and NLSP in this work are same as those of our previous work \cite{Explanation-6}, where gluino pair production was
used for the excess. This enables us to compare directly the two explanations.
\end{itemize}

\section{Z-peaked excess in our scenario} \label{sec:results}

\begin{figure}[thb]
\centering
\includegraphics[width=7.1cm]{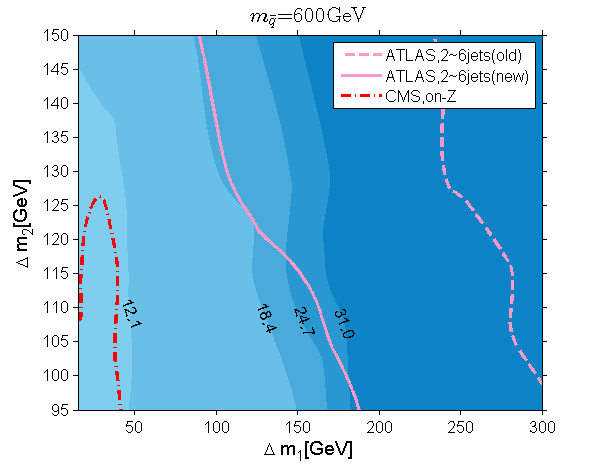} \hspace{0.1cm} \includegraphics[width=7.0cm]{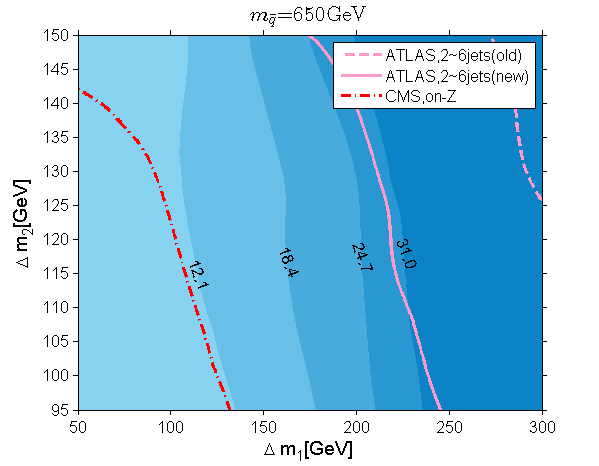}\\
\includegraphics[width=7.1cm]{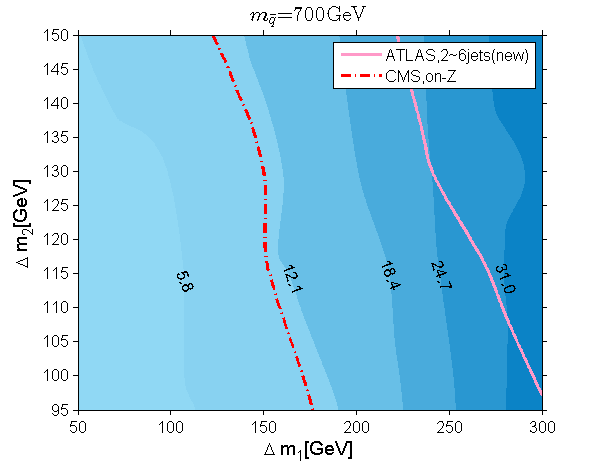} \hspace{0.1cm} \includegraphics[width=7.0cm]{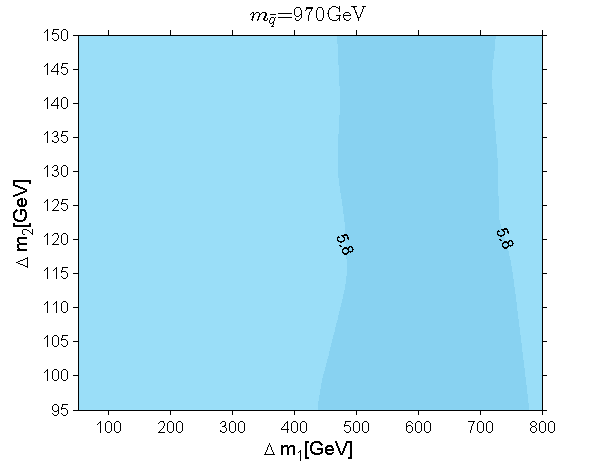}\\
\caption{ Constant contours of the event number for the ATLAS leptonic-$Z$ $ + jets + M_T^{Miss}$ signal on the $\Delta m_1 - \Delta m_2$
plane in heavy gluino case. For each $m_{\tilde{q}}$, the region between the contour marked by the number 12.1 and that by 24.7 can
explain the excess at $1\sigma$ level, and that between the lines marked
by 5.8 and 31 respectively can account for the excess at $2 \sigma$ level. The parameter spaces that coincide with different
SUSY searches are also presented, which are right bounded by different types of lines. The dotted line and solid line are the boundaries
coming from the ATLAS preliminary and updated searches for $2\sim6$ $jets + E_T^{miss}$  signal respectively,
and the dash dotted line is for the CMS constraint. For the case $m_{\tilde{q}}=700GeV$, the constraint
from the ATLAS preliminary search for $2\sim6$ jets + $E_T^{miss}$ signal is too weak to be drawn on the plane, and
for the case $m_{\tilde{q}}=970GeV$, there are actually no boundaries on the plane. }
\label{fig2}
\end{figure}

From the ATLAS analysis on the leptonic-$Z + jets + E_{T}^{miss} $ channel presented in \cite{ATLAS-Z-Excess}, one can infer that the
event number of the excess is $18.4 \pm 6.3$ after including the statistical and systematic uncertainties \cite{Explanation-6}. This
means that, if one wants to explain the excess at $1 \sigma$ and $2 \sigma$ levels, the SUSY signal number after cuts should satisfy
$12.1 \leq N_{ll}  \leq 24.7$ and $5.8 \leq N_{ll} \leq 31$ respectively. In order to find the parameter space that can produce
the required event number, we fix $m_{\tilde{g}} = 4.5 {\rm TeV}$ (heavy gluino case) and
$m_{\tilde{g}} = 1.5 {\rm TeV}$ (light gluino case) separately, and for each case, we perform a grid scan over the parameters $\Delta m_1$ and $\Delta m_2$ by choosing
a series of $m_{\tilde{q}}$.  For each parameter point, we calculate the squark pair production rate at the 8 TeV LHC by the package Prospino \cite{prospino}, and
generate the parton level events for the considered process with MG5\_aMC\cite{mg5}, which includes Pythia\cite{pythia} for parton showering and hadronization. Then we use
the package CheckMATE-1.2.0\cite{checkmate} which contains fine-tuned fast detector simulation code Delphes3.0.10\cite{delphes}
to repeat the analyses of various experiments.  These experiments include the ATLAS on-Z search\cite{ATLAS-Z-Excess}, the CMS on-Z search\cite{CMS-Leptonic-Z},
and the ATLAS $2\sim6$ jet +$E_T^{miss}$ searches\cite{ATLAS-6jet-new,ATLAS-6jet-old}, among which the first one is used to generate
the signal of the excess, and the other ones sever as constraints. In \cite{Explanation-6}, we encoded the cuts
for those experiments in the package CheckMATE-1.2.0, and the validation of them indicated that our calculations coincide with
the corresponding experimental analyses at $20\%$ level. In implementing the
constraints of the SUSY searches on the parameters, we define for each search the ratio
$R = max (N_{S,i}/ S_{obs,i}^{95\%})$, where $N_{S,i}$ is the event number of the SUSY signal in the $i$th SR of the search,
$S_{obs,i}^{95\%}$ is its $95\%$ upper limit usually provided in the experimental report, and the $max$ is over all SRs defined in the search. Obviously, only in case that $R < 1$,
the corresponding parameter point is experimentally allowed at $95\%$ C.L..

\begin{figure}[thb]
\centering
\includegraphics[width=7.1cm]{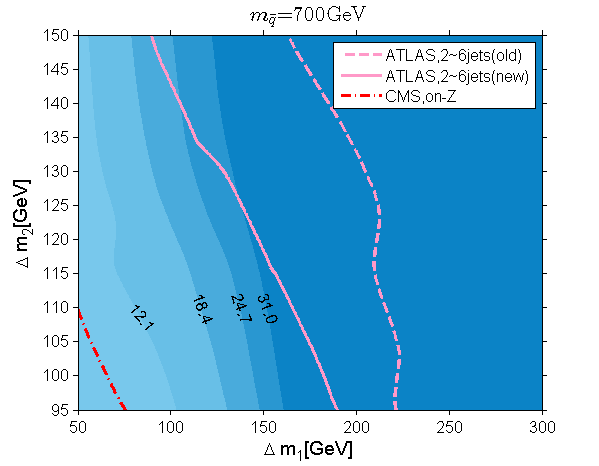} \hspace{0.1cm} \includegraphics[width=7.0cm]{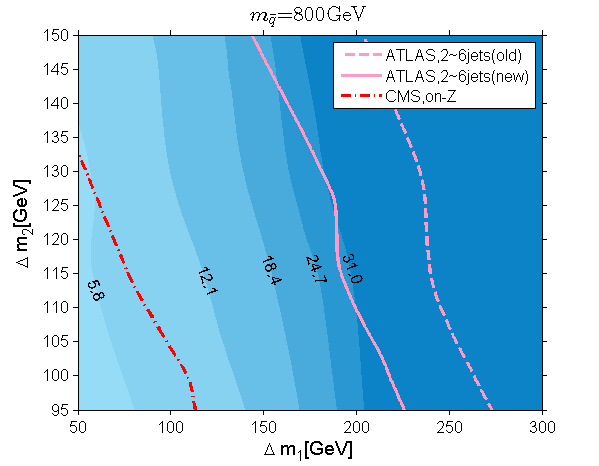}\\
\includegraphics[width=7.1cm]{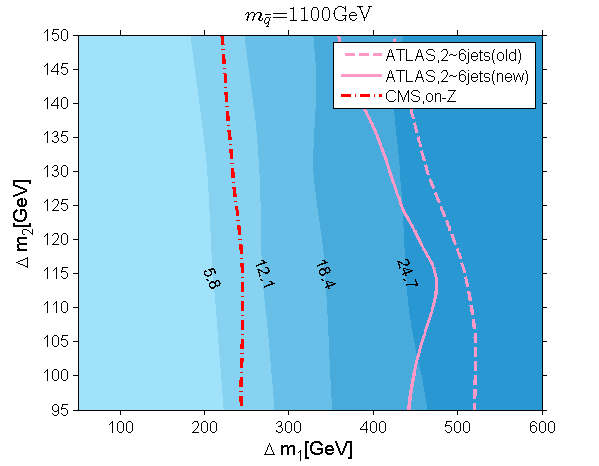} \hspace{0.1cm} \includegraphics[width=7.0cm]{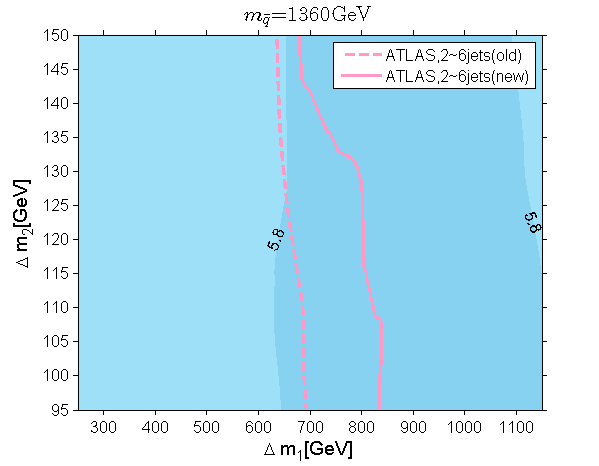}\\
\caption{Similar to Fig.\ref{fig2}, but for the light gluino case, $m_{\tilde{g}} = 1.5 {\rm TeV}$. Note that for
$m_{\tilde{q}}=1360 {\rm GeV}$, the CMS constraint is too weak to be drawn on the plane.}
\label{fig3}
\end{figure}

In Fig.\ref{fig2} and Fig.\ref{fig3}, we present on $\Delta m_1 - \Delta m_2$ plane the constant contours of the event number for
the ATLAS on-Z analysis in the heavy gluino and light gluino cases respectively. For each $m_{\tilde{q}}$, the region between the
contour marked by the number 12.1 and that by 24.7 can explain the excess at $1\sigma$ level, and that between the contours marked
by 5.8 and 31 respectively can account for the excess at $2 \sigma$ level. The parameter spaces that coincide with various
SUSY searches are also presented, which are right bounded by different types of lines (note that a compressed SUSY mass spectrum
is helpful to evade the LHC constraints). The dotted line and solid line are the boundaries
coming from the ATLAS preliminary and updated searches for $2\sim6$ $jets + E_T^{miss}$ signal
respectively, and they are obtained by setting the corresponding $R$ values at 1. The constraint from the CMS on-$Z$ search is obtained in a similar way,
and shown by dash dotted lines.

From Fig.\ref{fig2} for the heavy gluino case, one can learn following facts:
\begin{itemize}
\item For all the choices of $m_{\tilde{q}}$, the strongest constraint comes from the CMS dedicated on-Z counting experiment, and the weakest one is the ATLAS preliminary search for $2\sim6$ jets + $E_T^{miss}$ signal.
\item With the increase of the squark mass, $\Delta m_1$ is allowed to vary within a wider range. In this case, the improved cut efficiency due to the enlarged $\Delta m_1$ can compensate for the decrease of the squark pair production rate. As a result, even for $m_{\tilde{q}} \lesssim 800 {\rm GeV}$ the central value of the excess ($18.4$ events) can still be obtained if only the
    constraints from the ATLAS searches for the $jets + E_{T}^{miss}$ signal are considered.
\item In the CMS dedicated on-Z counting experiment, six signal regions discriminated by the jet number $n_{j}$ and $E_T^{miss}$ were considered (see Table 2 in \cite{Explanation-6}).
    We checked that for our scenario the tightest constraint of the experiment comes from $n_{j} \geq 2$ SRs with $E_{T}^{miss}$ either satisfying $200 {\rm GeV} \leq E_T^{miss} \leq 300 {\rm GeV}$ (called SR-II hereafter) or satisfying $E_T^{miss} > 300 {\rm GeV}$ (called SR-III hereafter)\footnote{In more detail, from our calculations we learn that the SR-II is more powerful than the SR-III in limiting our scenario for the region defined by $100 {\rm GeV} \lesssim \Delta m_1 \lesssim 150 {\rm GeV}$ and $95 {\rm GeV} \lesssim \Delta m_2 \lesssim 150 {\rm GeV}$ as well as that defined by $150 {\rm GeV} \lesssim \Delta m_1 \lesssim 200 {\rm GeV}$ and $95 {\rm GeV} \lesssim \Delta m_2 \lesssim 120 {\rm GeV}$ for all panels in Fig.\ref{fig2}. Consequently, the boundaries of the CMS experiment in the cases of $m_{\tilde{q}} = 650, 700 {\rm GeV}$ are mainly determined by the SR-II, while in the case of $m_{\tilde{q}} \gtrsim 750 {\rm GeV}$, they are determined by the SR-III.}. In either case, the signal of the ATLAS on-$Z$ search has a large overlap with that of the CMS on-$Z$ search, so due to the tension of the two search results the event number of the ATLAS experiment is always upper bounded by about 11 for $m_{\tilde{q}} \lesssim 750 {\rm GeV}$ after considering the CMS constraint. We also checked that, with the further increase of $m_{\tilde{q}}$ from about $750 {\rm GeV}$, the maximal reach of the event number drops either because the tension of the ATLAS and CMS data becomes stronger for moderately heavy $\tilde{q}$ or because the squark pair production rate is sufficiently suppressed for heavy $\tilde{q}$.

\item The lower right panel of Fig.\ref{fig2} indicates that there are actually no boundaries on the $\Delta m_1-\Delta m_2$ plane for $m_{\tilde{q}} = 970 {\rm GeV}$.
     In this case, the maximal reach of the event number is 6.8, which is still within the $2\sigma$ range of the excess.
\end{itemize}

\begin{table}[t]
\caption{Details about the constraints in light gluino case for two points $S_1$ and $S_2$, which are taken from the lower left and right panels in Fig.\ref{fig3} respectively, and
are defined by $S_1$: $m_{\tilde{q}} = 1100 {\rm GeV}, \Delta m_1 = 458 {\rm GeV}, \Delta m_2 = 105 {\rm GeV}$ and $S_2$: $m_{\tilde{q}} = 1360 {\rm GeV},
\Delta m_1 = 687 {\rm GeV}, \Delta m_2 = 105 {\rm GeV}$. In this table, $SR_{max}$ represents the SR which provides the strongest constraint
on the parameter for a certain experiment, and $\epsilon$ is the total cut efficiency of the SR. } \label{constraint-table}
\begin{tabular}{|c|c|c|c|c|c|c|c|c|c|c|c|c|c|}
\hline
\multirow{2}{*}{Point} & \multirow{2}{*}{$\sigma$(fb)} & \multicolumn{4}{c|}{CMS\cite{CMS-Leptonic-Z}}  & \multicolumn{4}{c|}{ATLAS(updated)\cite{ATLAS-6jet-new}} & \multicolumn{4}{c|}{ATLAS(preliminary)\cite{ATLAS-6jet-old}} \\ \cline{3-14}
 &  & $\text{SR}_{max}$ & ~~$\epsilon$~~ & $\text{S}_{obs}^{95\%}$ & ~~R~~ & $\text{SR}_{max}$ & ~~$\epsilon$~~ & $\text{S}_{obs}^{95\%}$ & ~~R~~ & $\text{SR}_{max}$ & ~~$\epsilon$~~ & $\text{S}_{obs}^{95\%}$ & ~~R~~ \\ \hline
$\text{S}_1$ & 32.9 & SR-III & 3.3\% & 7.6 & 2.77 & 2jm & 14.7\% & 90 & 1.09 & CM & 9.2\% & 81.2 & 0.75 \\ \hline
$\text{S}_2$ & 6.3 & SR-III & 4.2\% & 7.6 & 0.67 & 2jt & 21.0\% & 38 & 0.71 & BM & 12.4\% & 14.9 & 1.06 \\ \hline
\end{tabular}
\end{table}

Next we turn to the light gluino case.  From the event contours in Fig.\ref{fig3}, one can get following information:
\begin{itemize}
\item The dependence of the ATLAS event number on the parameters $\Delta m_1$ and $\Delta m_2$ is quite similar to that in the heavy gluino case, and so are
the dependencies of the constraints.
\item For squarks lighter than about $1200 {\rm GeV}$, the tightest constraint comes from the CMS experiment, which is similar to the heavy gluino case;
but with the increase of the squark mass, the strongest constraint may come from the ATLAS preliminary search for $jets + E_T^{miss}$ signal,
which is shown in the lower right panel of Fig.\ref{fig3}.  The underlying reason for such a feature is that the capabilities of the SRs defined in the
experiments in limiting SUSY depend on the configuration of the SUSY spectrum, and without the specification of the spectrum, there is no definite
conclusion about which is the strongest. To illustrate this, we take one point from each of the two lower panels in Fig.\ref{fig3} as an example, and show
the details of the constraints in Table \ref{constraint-table}. This table indicates that for the points $S_1$ and $S_2$, the tightest constraints of the ATLAS
preliminary search for $jets + E_T^{miss}$ are the SRs CM and BM respectively, and those of the ATLAS updated search correspond to the SRs 2jm and 2jt respectively.
For the point $S_1$, the constraint from the former experimental analysis is weaker, while for the point $S_2$, the situation reverses.

\item  We checked that the ATLAS excess can be explained at $2 \sigma$ level by squarks with mass up to about $1.4 {\rm TeV}$.
The wider mass range in comparison with the heavy gluino case is mainly due to the larger rate of the squark pair production
in light gluino case. We also checked that the central value of the excess can be achieved for $m_{\tilde{q}} \lesssim 1.2 {\rm TeV}$ if only the constraints from
the ATLAS searches for the $jets + E_{T}^{miss}$ signal are considered, and that if the constraint from the CMS experiment is further considered, the
maximal event number is only 9.5. The latter fact implies that the heavy gluino case is able to provide a slightly better explanation.

\item Note that for the lower right panel where $m_{\tilde{q}} = 1.36 {\rm TeV}$ and $m_{\tilde{g}} = 1.5 {\rm TeV}$, the effect from the squark-gluino associated production
on the LHC searches is non-negligible. Discussing such an effect is beyond the scope of this work.
\end{itemize}

Before we end this section, we have four comments about our explanation. First, comparing Fig.\ref{fig2} and Fig.\ref{fig3} in this work with Fig.2 in \cite{Explanation-6}
where gluino pair production with the decay mode $\tilde{g} \to q \bar{q} \tilde{\chi}_2^0 \to q \bar{q} Z \tilde{\chi}_1^0$ was used to explain the excess,
we conclude that the squark explanation allows for a significantly wider range of $m_{\tilde{q}}$ as well as a less compressed SUSY mass spectrum.
One underlying reason we think is that the cut efficiency of the ATLAS on-$Z$ search is usually larger for the squark explanation than for the gluino explanation,
and consequently, moderately heavy squarks are still capable of explaining the excess. Second, we emphasize again that
in our simple scenario, both the event number of the ATLAS signal and the constraints are determined by few SUSY parameters. Consequently, the capability of our scenario
to interpret the excess is limited. As we mentioned at the end of Sec.II, a more complex scenario with higher dimensional
SUSY parameters may improve this situation, but it needs tremendous calculation to search for the relevant parameter space\cite{Explanation-8}.
Third, we note that so far there have appeared other simple SUSY scenarios to explain the excess, and the studies of these scenarios
indicated that they can explain the excess at $1\sigma$ level in certain narrow SUSY parameter spaces \cite{Explanation-9,Explanation-10}.
This conclusion is slightly better than ours, where the best explanation is about $1.2 \sigma$ away from the central value of the excess.
Three factors may contribute to the difference:
\begin{itemize}
\item The difference in the theoretical hypothesis on SUSY, which determines the kinematical distributions of the SUSY signals. For example, both
\cite{Explanation-9} and \cite{Explanation-10} utilized the production $p p \to \tilde{g} \tilde{g}$ with the loop-induced decay $\tilde{g} \to g
\tilde{\chi}^0_i \to g Z \chi $ ($\chi$ denotes the lightest neutralino in \cite{Explanation-9} and the gravitino in \cite{Explanation-10})
to explain the excess. Comparing their interpretations with ours, one can learn that, although all of them
considered the two-body decay of a strongly produced SUSY particle, due to the difference of the properties for the parent sparticles such as
their spins and production channels, their kinematical distributions may differ greatly even when their predictions
on the event number of the excess are same. One can get this point by comparing the $E^{T}_{miss}$ distribution of the benchmark points $P_1$ and $P_2$ in this paper
with that of the best point in \cite{Explanation-10}, which are presented in Fig.\ref{fig4} of this work and in Fig.6 of \cite{Explanation-10}
respectively.  As a result of the difference, in principle there might exist SUSY points for which the ATLAS leptonic-$Z$ signal is moderately enhanced and
meanwhile the CMS signal is appropriately suppressed.

\item The uncertainties induced by related simulations. For all scenarios to explain the excess, simulations of the experimental searches for SUSY
have to be done. As shown in the appendices of \cite{Explanation-6} and \cite{Explanation-10} where the validations of the simulations were explicitly presented,
the uncertainties of the simulations are at the level of $20\%$, and therefore, the calculation performed by different groups may result in
a significant deviation. As far as our simulations are concerned, the computed efficiency for the ATLAS signal event is less than that presented by the ATLAS collaboration
for the selected SUSY point by about $10\%$, and by contrast our efficiency for the CMS search is slightly
larger than that in the CMS report.

\item The treatment of the CMS constraint. From the CMS report presented in \cite{CMS-Leptonic-Z}, one can only infer the approximate value of
the $S^{95}_{obs}$ for the SR-III. Confronted with such a situation, we in our previous work \cite{Explanation-6} calculated the $S^{95}_{obs}$s
for all the six SRs by the asymptotic $CL_s$ prescription \cite{CLs} (see Table 2 of \cite{Explanation-6}). Furthermore, we pointed out that the
SR-II may provide a stronger constraint on the parameters than the other SRs in discussing Fig.\ref{fig2} of this work. This conclusion
indicates that the calculation of all $S^{95}_{obs}$s is necessary; but on the hand, since the values
of the  $S^{95}_{obs}$s were not explicitly given in other previous literatures and meanwhile they depend on calculation method, there might exist
deviations for different authors in considering the CMS constraint. Let's show this point by an explicit example. During the revision of this manuscript,
the paper \cite{Li-Z-excess} appeared to interpret the excess in the NMSSM extension with a Dirac gluino, and the authors presented the details
about their calculation of $S^{95}_{obs}$s. Briefly speaking, the calculation in \cite{Li-Z-excess} differs from ours in at least two aspects. One is the authors of \cite{Li-Z-excess}
used the standard Bayesian procedure in the calculation, while we used the asymptotic $CL_s$ method \cite{CLs}. The other is that the work \cite{Li-Z-excess}
had considered the theoretical uncertainty in calculating the signal, and by contrast, we ignored such an effect. As a result of the differences,
the values of $S^{95}_{obs}$s in \cite{Li-Z-excess} are usually larger than our predictions by about $15\%$, and consequently, the CMS constraint
is significantly relaxed in \cite{Li-Z-excess}.

\end{itemize}
Finally, we note that in our scenario the singlino-dominated LSP is usually heavier than about $450 {\rm GeV}$, and one may wonder
how such a heavy dark matter (DM) achieves its measured relic density. In this case, the possible annihilation final states of the DM
include $f\bar{f}$, $VV$, $H_i H_j$, $A_i A_j$ and $H_i A_j$, where $f$ ($V$) denotes any of the fermions (vector bosons) in
the SM, and $H_i$ ($A_j$) denotes a CP-even (CP-odd) Higgs boson (see \cite{NMSSM} and references therein). The easiest way to achieve
the density is through the $s$-channel annihilations mediated by a singlet-dominated Higgs boson, where just like the light DM
case discussed in \cite{Cao-NMSSM-DM}, the Higgs boson mass as well as the self-coupling coefficient $\kappa$ for the
singlet fields play an important role in tuning the annihilation rate.

\section{Distributions of the excess} \label{sec:compare}

\begin{table}[t]
\centering
\begin{tabular}{|l|c|c|c|c|c|}
\hline
                                            & $P_1$     & $P_2$     & $P_3$     & SM        \\
\hline
$m_{\tilde{g}}$ (GeV)                       & 4500      & 4500      & 750       & -         \\
$m_{\tilde{q}}$ (GeV)                       & 650       & 700       & 4500      & -         \\
$m_{\tilde{\chi}_2^0}$ (GeV)                & 530       & 500       & 650       & -         \\
$m_{\tilde{\chi}_1^0}$ (GeV)                & 430       & 370       & 565       & -         \\
\hline
Events in ATLAS on-Z SR(8TeV,20.3$fb^{-1}$) & 11.0      & 18.2      & 10.5      & 10.6      \\
$\chi^2_{E_T^{miss}}(d.o.f.=9)$             & 7.6       & 5.6       & 8.0       & 14.8      \\
$\chi^2_{H_T}(d.o.f.=7)$                    & 5.7       & 2.6       & 6.0       & 13.8      \\
$\chi^2_{n_j}(d.o.f.=5)$                    & 6.9       & 4.7       & 6.8       & 14.1      \\
$\chi^2_{total}(d.o.f.=21)$                 & 20.2      & 12.9      & 20.7      & 42.7      \\
\hline
Events in ATLAS on-Z SR(14TeV,10$fb^{-1}$)  & 172.8     & 168.2     & 103.6     & 18.2      \\
Expected significance(14TeV,10$fb^{-1}$)    & 25        & 24        & 14        & -         \\
Expected significance(8TeV,20.3$fb^{-1}$)   & 2.4       & 4.0       & 2.3       & -         \\
\hline
\end{tabular}
\caption{Detailed information of the benchmark points $P_1$, $P_2$ and $P_3$, including their contributions to the ATLAS excess and behaviors
at the LHC-14.}
\label{table}
\end{table}

\begin{figure}[t]
\centering
\includegraphics[width=5.0cm,height=6.0cm]{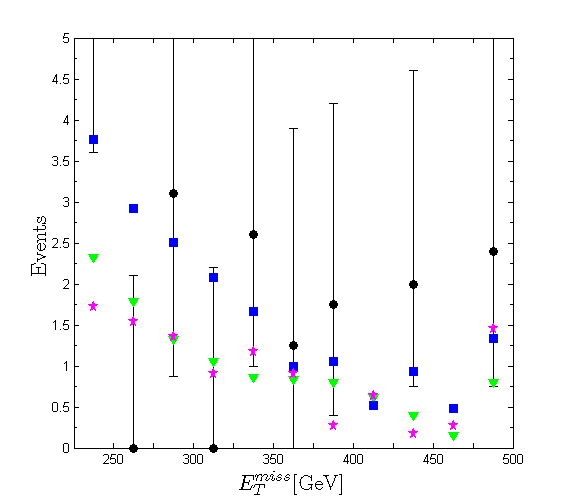} \includegraphics[width=5.0cm,height=6.0cm]{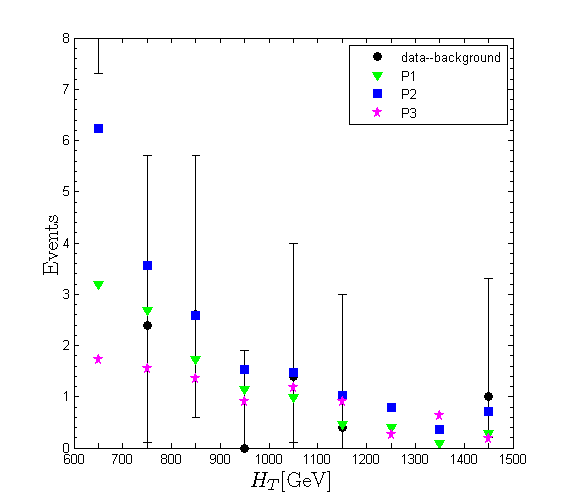}\includegraphics[width=5.0cm,height=6.0cm]{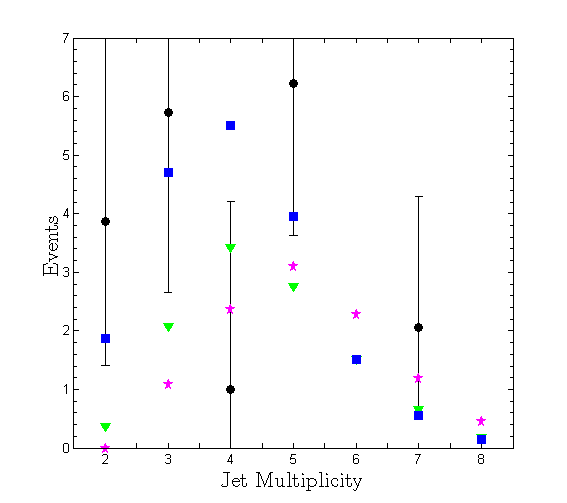}
\caption{Distributions of $E_T^{miss}$, $H_T$ and the jet multiplicity predicted by the three benchmark points. As a comparison, the experimental data for the
ATLAS on-$Z$ excess with the expected SM background extracted are also shown. Note that as did in \cite{Explanation-3},
only the statistical uncertainties are included in the error bars of the data.}
\label{fig4}
\end{figure}

In this section, we investigate whether our explanation can reproduce the distributions of the excess reported by the ATLAS collaboration. For this end, we concentrate on three
benchmark points P1, P2 and P3 with point P3 taken from our previous work \cite{Explanation-6}. Point P1 and P3 correspond to the best points after considering all the
constraints in the squark explanation and the gluino explanation respectively, and contribute 11 and 10.5 events to the excess.
By contrast, point P2 only satisfies the constraints from the ATLAS jets $ + E_T^{miss}$ searches, but it can reproduce the central value of the excess.
Detailed information of these points is presented in Table \ref{table}.

In order to compare our explanation with the experimental data for various distributions, we generate the distributions of $E_T^{miss}$, $H_T$ (the scalar sum of
the $P_T$s for the leptons and signal jets) and the jet multiplicity $n_j$ in the electron and muon combined channel for each parameter point. In getting
the distributions of $E_T^{miss}$ and $H_T$, we include the overflow events into the last bin.
The corresponding results are shown in Fig.\ref{fig4}, where the black solid circle with the error bar stands for the data obtained by the ATLAS experiment
with the expected SM background subtracted \cite{ATLAS-Z-Excess}, and the predictions of the points P1, P2 and P3 are marked by
triangles, squares and asterisks respectively.  To quantize the difference between the theoretical predictions and the corresponding
experimental data, we define a $\chi^2$ function for each distribution by a simple way
\begin{eqnarray}
\chi^2 = \sum_i \frac{(s_i - \hat{s}_i)^2}{(\delta s_i)^2},
\end{eqnarray}
where $s_i$ is the theoretical prediction in the $i$th bin, $\hat{s}_i$ is the corresponding experimental datum, and $\delta s_i $ is the error of the datum. In Table \ref{table},
we show the values of $\chi^2 $ for different distributions.

From Fig.\ref{fig4} and Table \ref{table}, one can learn that all points, especially point P2, can reproduce the distributions in an excellent way.
The only significant difference between the squark explanation and the gluino explanation in generating the distributions comes from the jet
multiplicity, i.e. for the former explanation, the $n_j$ distribution peaks at 3, while for the latter explanation, it peaks at 4.  Due to the large
errors of the data at current time, we can not determine which explanation is preferred to account for the excess.

\section{Test of our explanation at the 14 TeV LHC} \label{sec:test}

Considering that the squark pair production rate at the LHC-14 can be greatly enhanced in comparison with that at the LHC-8, one may expect that the squark
explanation will be tested very soon at the LHC. We investigate this issue by considering the lepton-$Z + jets + E_{T}^{miss}$ signal of
the production at the LHC-14. For simplicity, we assume the same cuts as those of the ATLAS on-$Z$ search at the LHC-8, and estimate the SM
background of the signal. In doing this, we suppose that the dominant background at the LHC-14 comes from the same processes as those at the
LHC-8, which include flavor-symmetric backgrounds, $Z$ + jets, rare top and diboson\cite{ATLAS-Z-Excess}. Since it is hard to
get accurate background events by directly simulating the processes at the LHC-14, we simulate each background process at the LHC-14 and LHC-8
separately to get the ratio of their rates after cuts, then we scale the background at the LHC-8, which was given in the ATLAS
report \cite{ATLAS-Z-Excess}, by this ratio. We realize that the results obtained in this way may deviate significantly
from their true values, but without any detailed information of the ATLAS detector at the LHC-14, our results may serve as a rough
estimate of the background.

Once we know the signal and the total background after the cuts, we can calculate the expected significance by following formulae
\begin{eqnarray}
S = \frac{N_s}{\sqrt{N_b + \epsilon^2 N_b^2}},
\end{eqnarray}
where $N_s$ and $N_b$ denote the event numbers for the signal and the background respectively, and the coefficient
$\epsilon$ parameterizes the effect induced by systematic errors. In our calculation, we set $\epsilon = 30\%$ for 14TeV LHC,
which was adopted at the LHC-8\cite{ATLAS-Z-Excess}.

Assuming $10 fb^{-1}$ integrated luminosity at the LHC-14, we present in Table \ref{table} the predictions of the points on the event
number of the signal as well as that of the background. This table indicates that the signals at the LHC-14 after cuts are enhanced by more than 9 times,
and by contrast the background is enhance by only about 2 times. As a result, either the squarks/gluino predicted by the points will be discovered,
or in case of the non-observation of the leptonic-$Z + jets + E_{T}^{miss}$ signal, the points will be excluded.
Moreover, comparing the squark explanation with the gluino explanation, we note that the former is more readily tested at 14TeV LHC.

\section{Conclusion} \label{sec:concl}

In this paper, we intended to explain the $3\sigma$ excess recently reported by the ATLAS collaboration in the search for the leptonic-$Z+jets+E_{T}^{miss}$ signal.
For this end, we considered the pair production of the first two generation squarks in the NMSSM with the decay chain
$\tilde{q} \to q \tilde{\chi}_2^0 \to q \tilde{\chi}_1^0 Z$. In order to maximize the signal rate and also to simplify our analysis,
we considered a singlino-dominated $\tilde{\chi}_1^0$ and a bino-dominated  $\tilde{\chi}_2^0$, and
assumed both $Br(\tilde{q} \to q \tilde{\chi}_2^0)$ and $Br(\tilde{\chi}_2^0 \to \tilde{\chi}_1^0 Z)$ at roughly $100\%$. With these assumptions, the parameters relevant to our analysis
include the common squark mass $m_{\tilde{q}}$, the gluino mass $m_{\tilde{g}}$ as well as the mass splittings $\Delta m_1 \equiv m_{\tilde{q}} -  m_{\tilde{\chi}_2^0}$ and
$\Delta m_2 \equiv  m_{\tilde{\chi}_2^0} - m_{\tilde{\chi}_1^0}$.

In order to find the parameter space that can explain the excess, we fixed $m_{\tilde{g}} = 4.5 {\rm TeV}$ and $m_{\tilde{g}} = 1.5 {\rm TeV}$ separately,
and for each case, we performed a grid scan over the parameters $\Delta m_1$ and $\Delta m_2$ by choosing a series of $m_{\tilde{q}}$.  Then for each parameter
point we encountered, we simulated the process $pp\to \tilde{q}\tilde{q} \to \tilde{\chi}_1^0 Z q \tilde{\chi}_1^0 Z q$ with the cuts adopted by the ATLAS on-Z
search, the CMS on-Z search, and the ATLAS $2\sim6$ jets +$E_T^{miss}$ searches respectively.
Based on our simulations, we have following conclusions:
\begin{itemize}
\item After considering the constraints from the ATLAS searches for $jets + E_{T}^{miss}$ signals, the central value of the ATLAS Z-peaked excess
can be obtained for $m_{\tilde{q}} \lesssim 1.2 {\rm TeV}$.
\item If the constraint from the CMS on-$Z$ search is further considered, more than 10 signal events are still attainable for $m_{\tilde{q}} \lesssim 750 {\rm GeV}$.
\item For the squarks as heavy as about $1.4 {\rm TeV}$, the squark pair production can still account for the excess at $2 \sigma$ level without conflicting with
any constraints.
\item Compared with the explanation by gluino pair production, the squark explanation allows for a significantly wider range of
$m_{\tilde{q}}$ as well as a less compressed SUSY mass spectrum.
\end{itemize}

Moreover, we also investigated whether the squark pair production can reproduce the distributions of the excess reported by the ATLAS collaboration. We found that,
quite similar to the gluino pair production, the squark pair production can fit the data quite well.  Probing the squark explanation at the 14 TeV LHC was also investigated,
and we concluded that only with $10 fb^{-1}$ integrated luminosity, the squarks that are able to explain the excess will be either discovered or excluded.

\section*{Acknowledgement}
We thank the authors of the CheckMate, especially Jamie Tattersall and Daniel Schmeier, for very useful discussion about the package. This
work was supported by the National Natural Science Foundation of China (NNSFC)
under grant No. 10821504, 11222548, 11121064, 11135003, 90103013 and 11275245,
and by the CAS Center for Excellence in Particle Physics (CCEPP).


\begin{thebibliography}{99}

\bibitem{Higgs-discovery}
  G.~Aad {\it et al.}  [ATLAS Collaboration],
  Phys.\ Lett.\ B {\bf 716}, 1 (2012)
  [arXiv:1207.7214 [hep-ex]];
   S.~Chatrchyan {\it et al.}  [CMS Collaboration],
  Phys.\ Lett.\ B {\bf 716}, 30 (2012)
  [arXiv:1207.7235 [hep-ex]].

\bibitem{ATLAS-Z-Excess}
  G.~Aad {\it et al.}  [ATLAS Collaboration],
  arXiv:1503.03290 [hep-ex].


\bibitem{Explanation-1}
  G.~Barenboim, J.~Bernabeu, V.~A.~Mitsou, E.~Romero, E.~Torro and O.~Vives,
  arXiv:1503.04184 [hep-ph].

\bibitem{Explanation-2}
  N.~Vignaroli,
  Phys.\ Rev.\ D {\bf 91}, no. 11, 115009 (2015)
  [arXiv:1504.01768 [hep-ph]].

\bibitem{Explanation-3}
  U.~Ellwanger,
  arXiv:1504.02244 [hep-ph].

\bibitem{Explanation-4}
  B.~Allanach, A.~Raklev and A.~Kvellestad,
  Phys.\ Rev.\ D {\bf 91}, 095016 (2015)
  [arXiv:1504.02752 [hep-ph]].

\bibitem{Explanation-5}
  A.~Kobakhidze, A.~Saavedra, L.~Wu and J.~M.~Yang,
  arXiv:1504.04390 [hep-ph].

\bibitem{Explanation-6}
  J.~Cao, L.~Shang, J.~M.~Yang and Y.~Zhang,
  JHEP {\bf 1506}, 152 (2015)
  [arXiv:1504.07869 [hep-ph]].

\bibitem{Explanation-7}
  B.~A.~Dobrescu,
  arXiv:1506.04435 [hep-ph].

\bibitem{Explanation-8}
  M.~Cahill-Rowley, J.~L.~Hewett, A.~Ismail and T.~G.~Rizzo,
  arXiv:1506.05799 [hep-ph].

\bibitem{Explanation-9}
  X.~Lu, S.~Shirai and T.~Terada,
  arXiv:1506.07161 [hep-ph].

\bibitem{Explanation-10}
  S.~P.~Liew, A.~Mariotti, K.~Mawatari, K.~Sakurai and M.~Vereecken,
  arXiv:1506.08803 [hep-ph].

\bibitem{Natural-SUSY}
  H.~Baer, V.~Barger and D.~Mickelson,
  Phys.\ Lett.\ B {\bf 726}, 330 (2013)
  [arXiv:1303.3816 [hep-ph]].

\bibitem{NMSSM}
  U.~Ellwanger, C.~Hugonie and A.~M.~Teixeira,
  Phys.\ Rept.\  {\bf 496}, 1 (2010)
  [arXiv:0910.1785 [hep-ph]].

\bibitem{Cao-NMSSM-DM}
  See, for example, J.~Cao, L.~Shang, P.~Wu, J.~M.~Yang and Y.~Zhang,
  arXiv:1506.06471 [hep-ph].

\bibitem{Franke:1995tc}
  F.~Franke and H.~Fraas,
  Int.\ J.\ Mod.\ Phys.\ A {\bf 12}, 479 (1997)
  [hep-ph/9512366].

\bibitem{prospino}
  W.~Beenakker, R.~Hopker and M.~Spira,
  hep-ph/9611232.

\bibitem{ATLAS-6jet-new}
  G.~Aad {\it et al.}  [ATLAS Collaboration],
  JHEP {\bf 1409}, 176 (2014)
  [arXiv:1405.7875 [hep-ex]].

\bibitem{ATLAS-6jet-old}
  The ATLAS collaboration,
  ATLAS-CONF-2013-047, ATLAS-COM-CONF-2013-049.

\bibitem{CMS-Leptonic-Z}
  V.~Khachatryan {\it et al.}  [CMS Collaboration],
  arXiv:1502.06031 [hep-ex].

\bibitem{Higgs-Mass-1}
  U.~Ellwanger,
  JHEP {\bf 1203}, 044 (2012)
  [arXiv:1112.3548 [hep-ph]].

\bibitem{Higgs-Mass-2}
  J.~ Cao {\it et al.},
  JHEP {\bf 1203}, 086 (2012)
  [arXiv:1202.5821 [hep-ph]].

\bibitem{Higgs-Mass-3}
  S.~F.~King, M.~Muhlleitner and R.~Nevzorov,
  Nucl.\ Phys.\ B {\bf 860}, 207 (2012)
  [arXiv:1201.2671 [hep-ph]].


\bibitem{mg5}
  J.~Alwall {\it et al.},
  JHEP {\bf 1106}, 128 (2011)
  [arXiv:1106.0522 [hep-ph]].

\bibitem{pythia}
  T.~Sjostrand, S.~Mrenna and P.~Z.~Skands,
  JHEP {\bf 0605}, 026 (2006)
  [hep-ph/0603175].

\bibitem{checkmate}
  M.~Drees {\it et al.},
  Comput.\ Phys.\ Commun.\  {\bf 187}, 227 (2014)
  [arXiv:1312.2591 [hep-ph]].
   J.~S.~Kim, D.~Schmeier, J.~Tattersall and K.~Rolbiecki,
  arXiv:1503.01123 [hep-ph].

\bibitem{delphes}
  J.~de Favereau {\it et al.}  [DELPHES 3 Collaboration],
  JHEP {\bf 1402}, 057 (2014)
  [arXiv:1307.6346 [hep-ex]].

\bibitem{CLs}
  A.~L.~Read,
  J.\ Phys.\ G {\bf 28}, 2693 (2002).

\bibitem{Li-Z-excess}
  R.~Ding, Y.~Fan, J.~Li, T.~Li and B.~Zhu,
  arXiv:1508.07452 [hep-ph].

\end{thebibliography}
\end{document}